\documentstyle[11pt]{article}

\begin{document}
\large

\def\lsim{\mathrel{\rlap{\lower3pt\hbox{\hskip0pt$\sim$}}
    \raise1pt\hbox{$<$}}}         
\def\gsim{\mathrel{\rlap{\lower4pt\hbox{\hskip1pt$\sim$}}
    \raise1pt\hbox{$>$}}}         
\def\dblint{\mathop{\rlap{\hbox{$\displaystyle\!\int\!\!\!\!\!\int$}}
    \hbox{$\bigcirc$}}}
\def\ut#1{$\underline{\smash{\vphantom{y}\hbox{#1}}}$}

\newcommand{\beq}{\begin{equation}}
\newcommand{\eeq}{\end{equation}}
\newcommand{\dem}{\Delta M_{\mbox{B-M}}}
\newcommand{\dega}{\Delta \Gamma_{\mbox{B-M}}}

\newcommand{\ind}[1]{_{\begin{small}\mbox{#1}\end{small}}}
\newcommand{\WA}{{\em WA}}
\newcommand{\SM}{Standard Model }
\newcommand{\QCD}{{\em QCD }}
\newcommand{\KM}{{\em KM }}

\newcommand{\appa}{\mbox{\ae}}
\newcommand{\CP}{{\em CP}}
\newcommand{\fy}{\varphi}
\newcommand{\hi}{\chi}
\newcommand{\al}{\alpha}
\newcommand{\as}{\alpha_s}
\newcommand{\gf}{\gamma_5}
\newcommand{\de}{\delta}
\renewcommand{\Im}{\mbox{Im}}
\renewcommand{\Re}{\mbox{Re}}
\newcommand{\GeV}{\mbox{ GeV }}
\newcommand{\MeV}{\mbox{ MeV }}
\newcommand{\matel}[3]{\langle #1|#2|#3\rangle}

\vspace{0.7cm}
\begin{flushright}
\large{
FERMILAB-PUB-92/158-T
\\ UND-HEP-92-BIG\hspace{0.1em}04
\\ TPI-MINN-92/30-T}
\\ June 1992
\end{flushright}
\vspace{0.4cm}
\begin{center} \LARGE {
Nonperturbative Corrections to Inclusive Beauty and Charm Decays : \hspace
{-.1em} \QCD versus Phenomenological
 Models} \end{center} \vspace{.4cm}
\begin{center} \Large I.I.Bigi\footnote{e-mail address:  BIGI@UNDHEP,
UNDHEP::BIGI}\\{\normalsize\it Dept.of Physics,
University of Notre Dame du
Lac, Notre Dame, IN 46556} \vspace{.4cm}
\\ N.G.Uraltsev\footnote{e-mail
address:  URALTSEV@UNDHEP ;  URALTSEV@LNPI.SPB.SU}
\\{\normalsize\it Dept.of
Physics,
University of Notre Dame du Lac, Notre Dame, IN 46556
\\{\normalsize and}\\
St.Petersburg Nuclear Physics Institute,
Gatchina, St.Petersburg 188350, Russia\footnote{permanent address}}
\vspace{.4cm}
\\ A.I.Vainshtein\footnote{e-mail address:
VAINSHTE@UMNACVX  } \\{\normalsize\it Theoretical Physics Institute,
University of Minnesota, Minneapolis, MN 55455
\\{\normalsize and}\\
G.I.Budker Nuclear Physics Institute,
Novosibirsk 630090, Russia}
\end{center}
\thispagestyle{empty} \vspace{.4cm}

\centerline{\Large\bf Abstract}
\vspace{.4cm}
We present a selfconsistent method for treating nonperturbative effects
in inclusive nonleptonic and semileptonic decays of heavy
flavour hadrons.
These effects give rise to powerlike corrections $\propto 1/m_Q^n\,$,
$n \ge 2$
with $m_Q$ denoting the heavy quark mass.
The leading correction to the semileptonic branching ratio
occurs for n=2. It is expressed in terms of the
vector-pseudoscalar mass splitting: $\delta BR\ind{sl}/BR\ind{sl}
\simeq BR\ind{nl}\, \cdot \,
6\,(\,(M_V^2-M_P^2)/m_Q^2)\cdot (c_+^2 - c_-^2)/2N_c$ and yields a
{\it reduction} of $BR\ind{sl}$. This nonperturbative correction
contributes to the nonleptonic
width with a sign opposite to that of the
perturbative terms that are non-leading in $1/N_c$.
In beauty decays the former
reduces the latter by 20 \%
whereas in charm decays they more or less cancel.
This leads to a {\it reduction} of $BR\ind{sl}$ by no more than
10 \% in
beauty decays and by a factor of roughly two in charm decays.
We confront these results with those obtained from
phenomenological models of heavy flavour decays
and find that such models are unable to mimic
these leading corrections by a specific choice of quark
masses or by invoking Fermi motion.
\newpage
\large
\addtocounter{footnote}{-4}

\hspace{-1em} Strong interactions affect weak decays of
hadrons in two ways:

(i)   They modify the {\it quark level} weak Lagrangian.

(ii)  They introduce bound state dynamics in the initial state and drive
hadronization in the final state.

It is in particular effect (ii) that we are at present unable to treat
in a satisfactory way. Yet it has always been expected that the
{\it inclusive} weak decays of hadrons containing a sufficiently heavy
quark Q can be dealt with quantitatively \cite {DU}.
The inclusive decay widths for heavy
flavour hadrons $H_Q$ is usually computed by considering the decay of a
`quasifree' quark Q in complete analogy to the decays of muons and $\tau $
leptons. Such a treatment is refined by computing the `ultraviolet'
renormalization of the effective $|\Delta Q|=1$ Lagrangian; this will
incorporate the main effects listed under point (i)
above. ( One can also include next-to-leading corrections due to
perturbative gluon bremsstrahlung.)
This decay mechanism is usually referred to as the
`spectator ansatz'. Among other things it predicts  uniform
1ifetimes and semileptonic branching ratios for hadrons of a
given flavour Q. This procedure
appears to work reasonably well in beauty decays. In charm decays on the
other hand the agreement of such expectations with the data
is much poorer as lifetime ratios of two to three emerge. This
shows the presence of large `preasymptotic'
corrections; yet it is fair to point out that even this
represents a vast improvement over the situation in strange decays.
Furthermore some ``non-spectator'' effects have been identified and were
found to be indeed of order unity in charm decays and thus in qualitative
agreement with the pattern in the data; the analogous corrections produce
only percent level effects in inclusive beauty decays
\cite {BU,VSbare,VSlog}.

There are however serious shortcomings to this simple approach:

$\bullet$ Bound state effects as well as hadronization -- listed above
under (ii)
 -- are not included in such a simple treatment. This severely
limits the accuracy level one can hope for.

$\bullet $ Various {\it phenomenological} decay models have
been employed to include these long distance forces. Yet they invariably
introduce new {\it a priori} free
input parameters that have to be fitted from the
data. This limits the predictive power of the theoretical treatment
considerably. Furthermore it is quite unclear how various systematic
uncertainties inherent in these models can be estimated in a reliable
way.

The goal of this letter is to outline in some detail a general procedure
that

(a)  enables us to calculate {\it inclusive} nonleptonic as well as
semileptonic decay rates in terms of the fundamental parameters
like the KM parameters;

(b)  allows for a {\it systematic} evaluation of the theoretical
uncertainties that can be refined {\it successively};

(c)  is {\it intrinsically} connected to \QCD without having to
call upon a phenomenological ``{\em deus ex machina}''.

The numbers we are going to present in this paper are intended to
illustrate the method and
show the trend of the effects. They should not be taken as the final
numerical
results. We will discuss how a consistent application of our approach
can lead to more precise numbers in the future.

The remainder of the paper will be organized as follows: in Sect.1 we
introduce the heavy mass expansion we are going to employ in our
analysis; in Sect.2 we discuss semileptonic branching ratios
before presenting our conclusions and an outlook in Sect.3.

\section{Heavy Mass Expansion}

In our analysis we will follow the general method outlined by
Shifman and Voloshin in ref.\cite{VSbare}. It was suggested there to
obtain the inclusive decay widths of heavy flavour hadrons $H_Q$ from the
absorptive parts of the forward amplitude $Q \rightarrow Q$ for the heavy
quark Q to second order in the weak Lagrangian; the simplest such
diagram is shown in Fig.1.  Integrating out the fields in the
intermediate state one can expand this absorptive part into a
series of inverse powers of the heavy
quark mass $m_Q$.

The leading contribution is obtained from
the absorptive part of the two-loop diagram in Fig.1: it is given by an
effective {\it operator} of the form

$$\hat{\Gamma}_o\;=\;\frac{G_F^2}{192\pi^3} \cdot N_c \cdot |KM|^2 \cdot
m_Q^5\,\bar{Q}Q \;\;\;.  \eqno(1) $$
$N_c$ denotes the number of colours and KM the appropriate KM
factors; the masses of $q_1$, $\bar q_2$ and $q_3$ have been ignored
here for simplicity.
The standard quasifree decay width
alluded to above is obtained when one evaluates the
{\it matrix element} of this operator between nonrelativistic
(with respect to $Q$) hadronic states
that contain the heavy quark Q \cite {VSbare,VSlog}; this is explained in
somewhat more detail later on.

When the intermediate
(anti)quark lines are cut and treated as {\it external} quark fields
-- see Fig.2 a,b -- one obtains contributions that are non-leading
in $1/m_Q$.
This procedure generates $SU(2)$ and $SU(3)$ {\it non-singlet}
operators that create
lifetime differences between the different kinds of mesons $H_Q$ with
a given
{\it heavy} flavour Q, namely $\bar Q u$, $\bar Q d$ and $\bar Q s$.
The diagram in Fig.2a [2b] is usually referred to as
`Weak Annihilation' or `W Exchange' [`Pauli Interference']; similar
diagrams also describe `Weak Scattering' in baryons. Attention in the
literature has so far focused on these kinds of preasymptotic effects
\cite{SONI,GUB,VSbare,VSlog,BIG}.

At tree level one obviously
obtains four-fermion operators of dimension {\it six} in this way
$$\hat{\Gamma}\ind{WA, PI} \propto \frac{G_F^2}{12\pi}\cdot |KM|^2 \cdot
m_Q^2 \:(\bar{Q}\Gamma Q)\, (\bar q\Gamma q)  \eqno(2) $$ where $q$ is
one of the light quark fields and $\Gamma$ denotes a combination of
$\gamma$-matrices.  These
operators thus yield a contribution of order $1/m_Q^3$ relative to the
spectator result stated in eq.(1). In ref.\cite{BU} this classification
was justified for {\it inclusive} transitions also in the
presence of gluon
emission and it was shown that the latter merely renormalizes -- in a
calculable way -- the coefficients of these operators.

The $1/m_Q$ expansion enables us to incorporate successively all
possible nonperturbative corrections to the
decay widths. There are also preasymptotic
contributions that affect {\it uniformly} all hadrons $H_Q$ of a
given heavy
flavour Q, at least within an isomultiplet; they lead actually to larger
corrections than those stated in eq.(2) as will be shown now.  These
contributions come from the quasifree two-loop amplitude of Fig.1 albeit
with the interaction of quarks with soft gluons included.
Those effects are
most conveniently dealt with by calculating the $Q\rightarrow Q$
amplitude
in the background gluon field: the inclusive width is then obtained as
an expansion in terms of {\it local} operators that contain $\bar QQ$ and
the gluon field. As the first byproduct one sees that there
cannot be terms
{\it linear} in $1/m_Q$.
For only operators of dimension {\it four} that contain $\bar{Q}Q$
could induce them.
Yet such an operator is either a total
derivative and thus has to vanish; or it can be reduced -- via the
equations
of motion -- to the original quasifree operator $\bar QQ$ thus merely
redefining the overall coefficient entering eq.(1).

The first nonleading operator in this
expansion carries dimension {\it five} and is the chromomagnetic dipole
operator for the heavy quark Q:
$$ D_G=\,\bar{Q}\,i\sigma_{\mu\nu}\hat{G}_{\mu\nu}\,Q \eqno(3) $$
where $\hat {G}_{\mu \nu} = g G_{\mu \nu}^a\lambda ^a/2$.
It is interesting to note that the same operator was used in
ref.\cite {BLOK2} to determine the leading corrections to the
factorizable amplitude for the {\it exclusive} decay
$B\rightarrow D\pi$.

As already stated there are dimension {\it six} four-fermion
operators that appear explicitly by cutting one of the light quark
propagators. We would like to add here that dimension six operators
that do not contain light quark fields as {\it external} legs (and hence
cannot cause a splitting among the decay widths of
hadrons within the same
isomultiplet) are in general of little practical relevance.  For such
operators can arise only by embedding the light quark lines into a vacuum
condensate.  Yet for purely left-handed weak vertices such effects are
proportional to the current quark masses; thus they will be small and in
general insignificant (unless one studies the
 question of $SU(3)$ breaking).

Operators containing additional gluon fields appear on the dimension
{\it seven} level and thus can give
corrections of order $1/m_Q^4$ only. Calculating them is in principle
straightforward, yet probably of little practical value since their
matrix elements are mostly unknown. Furthermore there are quite a
few of them and
for this reason it is unlikely that they can be directly related to
observable quantities. These contributions could therefore well represent
the intrinsic limit on the numerical  accuracy of our approach. Yet
since they are of order $1/m_Q^4$ they are expected to induce only rather
small corrections.  A presumably conservative estimate of the accuracy of
the expansion described here could be obtained by considering only the
factorizable contributions which
are expected to be dominant. It
should be possible to estimate their impact since there are fewer
operators that
contain them and these contributions have a simpler structure.

Once the {\it operator product expansion} has been obtained up to a
certain
order one has to address the delicate question of how to evaluate the
{\it matrix elements} of the various local operators.
For even the quasifree
operator $\bar QQ$ receives nonperturbative contributions.  This is most
clearly seen by again employing a heavy quark expansion.
The situation here
is actually simpler than the one encountered in a general application
of Effective Heavy Quark Theory \cite{VSheavy} (hereafter referred
to as {\em EHQT}):  for we are here
interested only in `zero recoil' amplitudes.
We can then simply use a nonrelativistic expansion
$$\bar{Q}Q = v_\mu \,
\bar{Q}\gamma_\mu Q - 1/4m_Q^2 \,\bar{Q}\,i\sigma G \,Q - 1/2m_Q^2
\,\bar{Q}\,(D^2-(v_\mu D_\mu)^2)\, Q \:+\:O(1/m_Q^3)\;\;\; \eqno(4)$$
As before, linear terms in $1/m_Q$ vanish due to the
equation of motion.

Concerning the matrix elements three observations can be made:

\hspace{-1em}(i) The first operator on the right hand side of eq.(4) is
the generator of the conserved charge that is associated with the heavy
flavour Q. Its matrix element taken between heavy flavour hadrons $H_Q$
therefore yields {\em exactly} unity.

\hspace{-1em}(ii) The matrix
elements of the second operator between meson states can be directly
expressed in terms of the mass splitting between the vector and the
pseudoscalar $\bar Qq$ boundstates; it is thus extracted from the
data. The
matrix element between the heavy flavour baryon $\Lambda _Q$ on the other
hand vanishes.

\hspace{-1em}(iii) The third term represents the kinetic
energy of the heavy quark in the presence of the
gluon background field. Its matrix element is quite possibly
different when taken between baryon rather than meson states.
So far we have not found a reliable way for extracting the size of these
matrix elements from a
direct phenomenological analysis. This operator enters the subleading
$1/m_Q$ corrections to the
masses of hadrons $H_Q$:
$$M_{H_Q}=m_Q+\mu_o+\mu_1^2/m_Q+\ldots\;\;\;\;.$$
Therefore one piece of information can be gained here by considering the
masses of charm and beauty hadrons and actually that combination
from which
the chromomagnetic contribution drops out:
$$(3M_D+M_{D^*}-4M_{\Lambda_c})-(3M_B+M_{B^*}-4M_{\Lambda_b}) \simeq $$
$$\simeq
2(1/M_D - 1/M_B) \cdot (\matel{\mbox {baryon}}{\bar{Q}\vec{D}^2 Q}
{\mbox {baryon}} - \matel{\mbox {meson}}{\bar{Q}\vec{D}^2 Q}
{\mbox {meson}})\;\;.\eqno(5)$$
(The vanishing of the anomalous dimension for the kinetic energy
has been taken
into account here). Once a precise value for the $\Lambda _b$ mass has
been obtained, we can then extract one combination of matrix
elements for the
$\bar{Q}\vec{D}^2 Q$ operator; the typical scale here however is probably
only about $50\MeV.$  We believe that an analysis based on \QCD sum rules
could be developed to estimate this term both in mesons and in baryons or
at least a complementary combination of matrix elements.

Fortunately many interesting quantities do not depend on this operator.
For it appears only in the nonrelativistic expansion of the
 operator
$\bar QQ$  and as such contributes with the same weight to all decays of a
given hadron $H_Q$; accordingly it will not affect the semileptonic {\it
branching ratio} which will be discussed next.

\section{Semileptonic Branching Ratio}

As a topical application of this general approach let us consider the
semileptonic branching fraction for beauty hadrons. As explained before
the $1/m_Q^2$ nonperturbative corrections to this quantity are given
by the matrix element of the chromomagnetic operator. A rather
straightforward calculation yields for nonleptonic transitions in the
external gluon field (hereafter we omit the obvious KM factors)
$$\hat{\Gamma}\ind{nl} = \frac{G_F^2 m_b^5}{192\pi^3}\cdot N_c\:
\{A_0 z_0 \cdot (\bar{b}\, b\, -\, \frac{1}{m_b^2}\,
\bar{b}\, i\sigma G \,b )\: -\:  A_2 z_2\cdot \frac{4}{m_b^2} \,\bar{b}\,
i\sigma G \,b ) \} \eqno (6) $$
where $A_{0}$, $A_{2}$ denote colour factors
$$A_0= c_1^2+c_2^2+\frac{2}{N_c}c_1c_2 + O(\as(m_b^2))\:,\; \; \;
A_2=\frac{2}{N_c}c_1c_2 + O(\as(m_b^2))\eqno(7)$$
$$ c_1=(c_+ +c_-)/2\:, \; \; \; c_2=(c_+ - c_-)/2\eqno(8) $$
and $z_0(m_c^2/m_b^2)$, $z_2(m_c^2/m_b^2)$ represent phase space factors
that reflect the sizeable mass of the $c$ quark:
$$ z_0 (x)=1-8x+8x^3-x^4-12x^2\log x\:,\; \; \; \; \;
z_2(x)=(1-x)^3\;\;\;.\eqno(9)$$
Obviously $z_0(0)=z_2(0)=1$ holds.

As stated before these calculations are most conveniently performed by
using the explicit expressions for the intermediate quark propagators
in the presence of a gluon background field. Employing specifically the
Fock-Schwinger (fixed point) gauge for the gluon field
(see ref.\cite {NVZS}
for details)
further facilitates such computations. The operator that emerges there is
actually $p_{\mu}\tilde {G}_{\mu \nu}\gamma _{\nu}\gf$; this
expression is
reduced to the chromomagnetic operator by applying the equations of
motion
to the $b$ field (or by confining oneself to nonrelativistic
beauty fields).

The two terms in eq.(6) with coefficients $A_0$ and $A_2$ show that there
are two distinct sources for the dimension five operator $D_G$:

\hspace{-1em}(a) The first emerges from the quasifree diagram of Fig.1
upon rewriting $\bar b \, (i\partial \! \! \! \backslash )^5 \, b$
in terms of
covariant derivatives $\bar b \, (i{\cal D} \! \! \! \backslash )^5 \, b$
that are simplified due to equation of motion; the $D_G$ operator is then
induced by Dirac commutators and thus reflects the conservation of
colour.

\hspace{-1em}(b) The second term represents
the chromomagnetic interaction of the various quarks in the internal
lines in Fig.1 with the external gluon field. It can
actually be proven in general that only the antiquark, but not the
quarks can contribute here when both weak vertices are purely left-handed.

For semileptonic transitions on the other hand one obtains
$$\hat{\Gamma}\ind{sl} = \frac{G_F^2 m_b^5}{192\pi^3}
\cdot z_0 \cdot (\bar{b}\, b
 - \frac{1}{m_b^2}\,
\bar{b}\, i\sigma G\, b ) \eqno (10) $$
{\em i.e.} there is no overall colour factor $N_c$, $c_+=c_-=1$
applies here
and the second term on the right hand side of eq.(6) is absent.

It is thus the operator $\bar{b}\, i\sigma G \,b$ that can generate
a shift in the predicted semileptonic branching ratio. Its matrix element
between the $\Lambda _b$ baryon vanishes. For B mesons on the other hand
it does not vanish; it can actually be expressed in terms of the mass
splitting between $B$ and $B^*$ mesons according to the following
relations:

$$  \, \matel{B}{\,\bar{b}\,i\sigma G\, b\,}{B} = -3
\matel{B^*}{\,\bar{b}\,i\sigma G \,b\,}{B^* } \eqno(11)$$
$$(M_{B^*}^2-M_B^2)\matel{B}{\,\bar{b}\, b\,}{B} = -1/2 \,
(\matel{B^*}{\,\bar{b}\,i\sigma G \,b\,}{B^*} - \matel{B}{\,\bar{b}\,
i\sigma G \,b\,}{B}) \eqno(12)$$
(it should be noted that eq.(12) does
not depend on the normalization of the operators and states.) The
modification of the nonleptonic width is then given by the
following factor:

$$r\: \simeq \:1\,-\,6\,\frac{M_{B^*}^2-M_B^2 }{m_b^2}
\cdot \frac{A_2}{A_0}\cdot\frac{z_2}{z_0}\;\;\;\;. \eqno(13)$$
Since the colour factor $c_2$ and thus also $A_2$ are negative this
nonperturbative correction actually enhances $\Gamma _{nl}$ and thus
{\it decreases} the semileptonic branching fraction.
The scale for the reduction of $BR\ind{sl}(B)$ is set
by the factor $6(M_{B^*}^2-M_B^2)/m_b^2 \simeq 15\%$;
that would translate to a reduction in $BR\ind{sl}(B)$ by about 1.5
percentage points, {\em i.e.} from, say, 12-13\% (see {\em e.g.}
ref.\cite{AP}) to 10.5-11.5\% -- if the remaining factor
$A_2 z_2/ A_0 z_0$
equaled unity.

The `kinematic' ratio $z_2/z_0$ must actually exceed unity as seen
from the following semi-quantitative argument: consider the limiting
case when the mass
of the charm quark approaches that of the beauty quark;
the mass scale that
in this case enters into the denominator of the chromomagnetic
term will be the
energy release $m_b - m_c$ rather than $m_b$.
Numerically we obtain from eq.(9):
$z_2/z_0 \simeq  1.5 $.
The ratio of colour factors $A_2/A_0$ on the other hand will reduce the
size of the effect. For the antiquark can interact with the external
gluon
field in the linear approximation only if the colour flow through the two
weak vertices is different. This leads to
a reduction factor $1/N_c$
and also makes the effect proportional to the coefficient $c_2$. Its size
is small in beauty decays and one obtains on the leading {\em log} level
$A_2/A_0\, \simeq \,-(0.15-0.2)$.
Combining all these factors we obtain
$$ \delta BR_{sl}/ BR_{sl} \simeq BR_{nl}\, \cdot \, (1 - r)
\simeq 3-4 \% $$
It is quite conceivable however that the subleading perturbative
corrections --  namely those due to the emission of hard gluons with
$|{\bf \vec{k}}|\sim m_b\,$ -- will increase this ratio
further and thus soften colour suppression. For the presence of such
gluons
immediately invalidates both reasons for the suppression
of the `leading' correction: it enables
all fields in the loops to contribute to the operator $D_G$ and
it can change the
unfavorable colour flow. Such effects are only of order
$\as$
and cannot bring any {\em log} enhancement.
Therefore it is unlikely
that they can completely eliminate the suppression; nevertheless
it is conceivable
that the next-to-leading perturbative correction to
the coefficient function may essentially change the numerical result.

Two general remarks are in order here:

\hspace{-1em} (i) It
would not invalidate a perturbative treatment of the problem at hand if
these higher order corrections indeed turned out to be larger than the
lowest order one:  for the suppression of the lowest order
result is due to the specific structure of the weak interactions, namely
their purely chiral nature, and has nothing to do with the strong forces.

\hspace{-1em} (ii) Since the mass difference $m_b-m_c$ is still large
compared to typical hadronic scales one can calculate the
coefficient of the
chromomagnetic operator $D_G$ within perturbation theory;
such a computation
thus presents only a {\em technical} challenge, but not one of a
principal nature.

It is quite intriguing to extrapolate this analysis down to the case
of charm
decays: since $6(M_{D^*}^2-M_D^2)/m_c^2 \sim 2$ and colour suppression is
less severe there
we find that a very large nonperturbative reduction arises here
for the semileptonic branching ratio of charm mesons.
In addition the dimension six four-fermion operators produce a large
decrease in the width
of $D^\pm$ through Pauli Interference \cite {BU,VSlog}.
This yields the
correct pattern as
observed by experiment, however definite quantitative predictions
seem to be
unreliable here.

There is another semi-quantitative observation that should be noted:
the nonperturbative term in eq.(6) that is responsible for increasing the
nonleptonic B decay width is -- through $A_2$ -- suppressed by $1/N_c$;
there is also a $1/N_c$ term -- due to $A_0$ -- appearing in the first
quasifree term
and it enters with the {\it opposite} sign there. Thus the two
terms that are
non-leading in $1/N_c$ tend to cancel each other. This provides another
illustration of preasymptotic \QCD corrections tending to
{\it dynamically}
suppress non-leading corrections in $1/N_c\;$ \cite{BU,BLOK,BLOK2}
and thus
to imitate the phenomenological prescription of retaining only the
leading
terms \cite{BUR}. It also shows on the other hand that such a recipe {\it
cannot} be viewed as a fundamental and universal rule. This
cancellation is
relatively mild for $\Gamma\ind{nl}(B)$ -- say up to 20\%; for the
nonleptonic width of $D$ mesons on the other hand it could
be complete. The situation here differs from what was found in
ref.\cite{BU} for flavour-dependent effects: there the cancellation
was due
to {\em perturbative} corrections and for this reason it was strong
both in
charm and in beauty; in the present case the corrections are
nonperturbative
and their role decreases significantly from charm to beauty.
It should also be noted that the authors of ref.\cite {BLOK2} found
that the terms that are non-leading in $1/N_c$ cancel almost
completely for the exclusive mode $B\rightarrow D\pi$; as we have seen
here this is not the case for the inclusive width.

Similar considerations can be applied to the inclusive width for
$b\rightarrow c \bar c s$ transitions. The relative weight of the
nonperturbative corrections could be quite sizeable there due to the small
amount of energy release that is available. Of course the overall size of
this width is also suppressed by phase space. We can also expect here even
more important modifications to the corresponding colour factor $A_2$
in eq.(7) coming from the subleading perturbative corrections than we
discussed for $b\rightarrow c \bar u d$. Yet these are `just'
technical problems that can and will be addressed in future work.
Lastly forthcoming experimental studies will allow the reliable
isolation of
this class of transitions.

The approach presented here can also be applied to the lepton
{\it spectra}
in semileptonic beauty decays.  The relevant operator expansion is then
given in terms of inverse powers of $(p_b-p_l)^2$ with $p_b\,[p_l]$
denoting
the momentum of the $b$ quark [lepton].  Thus the series blows up
near the endpoint in the lepton energy spectrum. Yet there
one can rely on the
results obtained from the {\em EHQT} approach. Since
$ m_b > m_c \gg \mu\ind{had}\;, \; \; m_b-m_c\gg \mu\ind{had}$
hold there exists a regime for the lepton energy where both expansions
are
valid simultaneously. This argument is actually rather similar to the
discussion of ref.\cite{GF}.

To summarize:
we have identified a correction to the inclusive nonleptonic decay
width that has no counterpart in the semileptonic width.
It may account for up to 15\% of the total
beauty width. The lowest loop estimate though yields
only about 3\%; this is roughly
similar in size to the
effect of interference coming from four-fermion operators of
dimension six. That a dimension five and a dimension six operator
have a similar impact here
is due to two factors: the colour
suppression of the dimension five
operator as well as the fact that it appears in the
two loop diagram and
thus contains an extra factor of
$1/\pi^2$. Basically
the same reason \cite{GUB} enhances the weight of the
standard corrections to the lifetimes. A
more exact estimate requires calculation
of the $O(\as)$ corrections; still it seems unlikely
that in the \SM it
could by itself shift the
semileptonic fraction by more than 1\% per
lepton flavour.

\section {Summary and Outlook}

We have outlined here a general method that allows to calculate the
inclusive
transition rates for the weak decays of heavy flavour hadrons.
It consists of four elements:

\hspace{-1em} (i) The forward amplitude $Q\rightarrow Q$ (more exactly
its absorptive part) is expanded into a series of local operators of
increasing dimension whose coefficients are proportional to powers of
$1/m_Q$. This operator expansion depends on the intermediate state,
namely whether one is considering nonleptonic or semileptonic
transitions, {\it i.e.}
$Q\rightarrow q_1\bar q_2 q_3\rightarrow Q$ or
$Q\rightarrow q_1\bar \nu l\rightarrow Q$; it is also
sensitive to the masses of the quarks $q_i$. On the
other hand it is universal for all hadrons carrying the heavy flavour Q.

\hspace{-1em} (ii) A nonrelativistic expansion is given for these local
operators, again in powers of $1/m_Q$.

\hspace{-1em} (iii)  The inclusive decay rate is obtained from the
matrix element of this operator expansion taken between the decaying
meson or baryon state.

\hspace{-1em} (iv)  The matrix elements for the operators that appear in
this final expansion are determined by symmetry arguments and/or by
relating
them to other observables like the masses of heavy flavour hadrons.
The size of these matrix elements in general depends on the type of
$H_Q$, {\it i.e.}
whether it is a meson or a baryon, whether it is charged
or not, whether it carries strangeness etc.

Our approach allows to incorporate nonperturbative effects in a
selfconsistent
way as corrections in a $1/m_Q$ expansion. We find:

$\bullet$ There are no corrections to the quasifree picture of order
$1/m_Q$.

$\bullet$ The leading nonperturbative corrections arise on the $1/m_Q^2$
level.  They are $SU(3)\ind{fl}$ invariant, {\em i.e.} affect the heavy
flavour meson decays in a uniform way independent of the flavour of the
light antiquarks. They enhance the nonleptonic decay width in mesons and
lead to a corresponding {\it reduction} in the semileptonic
branching ratio.

$\bullet $ The nonperturbative corrections that appear on the $1/m_Q^3$
level are not $SU(2)$ and $SU(3)$ invariant; thus they generate
differences
in the lifetimes and semileptonic branching ratios among all heavy
flavoured
hadrons.

$\bullet $ Due to the powerlike scaling behaviour in $1/m_Q$ all these
preasymptotic effects are much larger
in charm than in beauty decays.

$\bullet $ We have found some cases where nonperturbative and perturbative
corrections that are non-leading in $N_c$ contribute with the opposite
sign and thus tend to cancel each other. This provides a dynamical
explanation for some of the successes of the phenomenological prescription
to drop terms that are non-leading in $N_c$. Yet it also shows that such
a procedure cannot be expected to be of universal validity.

These findings can be stated in a more quantitative manner:

\hspace{-.8em} (1) As previously shown \cite {BU}
lifetime differences in B meson
decays cannot be expected to exceed the percent level while
lifetime ratios
of two emerge naturally for D meson decays.

\hspace{-.8em} (2) Nonperturbative corrections may reduce the
semileptonic branching ratio {\it uniformly} in B decays by up to 10\%
relative to the value inferred from the naive spectator ansatz. For D
decays they could well reduce it by a factor of two!

Point (2) is quite intriguing when comparing our approach with the
usual one
that involves modifying the quasifree result with phenomenological models.
The one class of parameters that is quite uncertain there is the mass for
the final state quarks, {\em i.e.} whether the small `current'
values should
be adopted or the higher `constituent' values.  Yet increasing the final
state quark masses over their `current' values will necessarily
{\it enhance}
the semileptonic branching ratio!  The \QCD approach presented here
on the other hand
demands that
it is the current masses that must be used in the
calculations.  Powerlike
nonperturbative corrections for the colourless quark-antiquark loop
itself
in Fig.1 are known to {\em increase} the hadronic width.
Some of these corrections could indeed effectively
mimic the insertion of
constituent masses into the  propagators; yet
being of order $1/m_b^6$ they are in fact strongly suppressed.
The corrections to the quark
loop discussed in ref.\cite{VSbare} for charm decays are subleading
contributions
resulting in terms that start with $1/m_b^4$. The leading
effects are actually
due to the interaction of the decay quarks with the light degrees
of freedom present in the initial hadron. They induce $1/m_Q^2$
corrections
that are explicitly calculable and in principle depend on the
nature of the
spectator (but not on its flavour).

We have outlined here a general procedure. It will be improved and
extended
in the future in four respects:

\hspace{-.5em} (i) Some of the numerical
predictions stated above were somewhat tentative since not all the
relevant
calculations have been performed yet. Since the `missing' computations
involve perturbation theory this presents just a technical delay and
not a stumbling block in principle.

\hspace{-.5em} (ii) The
real accuracy that can be obtained in
this approach is to be determined by
calculating terms of order
$1/m_Q^4$ and estimating the size of the relevant matrix
elements.

\hspace{-.5em} (iii) The case of heavy flavour baryons can be
and will be incorporated in a systematic way \cite {BU2}.

\hspace{-.5em} (iv) A
natural approach to the problem of $SU(3)\ind{fl}$
breaking in heavy flavour decays emerges from our treatment. It can
be expected quite generally
from the Heavy Quark Expansion outlined above that
the apparent size of $SU(3)\ind{fl}$ breaking scales like
$m_s/\mu\ind{had} \cdot 1/m_Q^2$; in other words $SU(3)\ind{fl}$ breaking
is expected to amount to no more than a few percent effects in beauty
decays \cite{BU2} -- whereas a conventional application
of $SU(3)$ arguments can do no better than allow for the
usual 20-30\% breaking effects.

\vspace{0.5cm}

{\bf ACKNOWLEDGEMENTS:} \hspace{.4em} N.U. and A.V.
gratefully acknowledge
illuminating discussions on the subject of this paper with
M.A.Shifman and
M.B.Voloshin.
N.U. thanks also V.M.Braun, M.Neubert and M.B.Wise for
informative discussions on {\em EHQT}. He has
benefitted from the inspiring atmosphere at the Theoretical Physics
Institute of the University of Minnesota; together with I.B.
he is grateful
to the members of the FNAL Theory Division for their kind hospitality
and in
particular to Joanne Cohn for her great tolerance and patience. This
work
was supported in part by the National Science Foundation under
grant number
PHY 89-09929.


\vspace{0.4cm}
\LARGE
{\bf Figure Captions}
 \vspace{0.4cm} \\
\large
{\bf Fig.1} \hspace{.5em} Simplest two-loop diagram describing the total
decay width $Q\rightarrow q_1\; \bar q_2\; q_3$ \vspace{.6cm} \\
{\bf Fig.2} \hspace{.5em} Diagrams representing flavour-dependent
corrections to widths:

\hspace{1.3em}{\bf a.} \hspace{-.1em} ``Weak Annihilation'' in heavy
meson
decays

\hspace{1.3em}{\bf b.} \hspace{-.1em} ``Pauli Interference'' effects
in heavy
meson decays

\end{document}